\pdfoutput=1
\documentclass[11pt]{article}
\usepackage{amssymb}
\usepackage{amsfonts}
\usepackage[activate={true,nocompatibility}, tracking=true]{microtype}
\usepackage{amstext,theorem}
\usepackage{amsopn}
\usepackage{amsmath,shadow,fancybox,graphics, amsmath}
\usepackage{epsf,color,url,siunitx}
\usepackage[T1]{fontenc}
\usepackage[utf8]{inputenc}
\usepackage[setpagesize=false]{hyperref}

\usepackage{geometry}
\usepackage[authordate,useprefix=true,footmarkoff,backend=biber,hyperref=true,backref=true,backrefstyle=all+,sorting=nyt,bibencoding=utf8,url=false,cmslos=true,maxbibnames=20,minbibnames=7]{biblatex-chicago}

\DeclareLanguageMapping{american}{cms-american}
\newcommand\nptextcite[1]{\textcite{#1}}

\begin{document}
\title{From physics to biology by extending criticality and symmetry breakings\footnote{Published as: \fullcite{longo2011c}}}

\author{Giuseppe Longo\footnote{Informatique, CNRS – ENS and CREA, Paris, longo@di.ens.fr, \url{http://www.di.ens.fr/users/longo}} , Maël Montévil\footnote{Informatique, ENS and ED Frontières du vivant,  Paris V, Paris }}
\maketitle

\tableofcontents

\subsection*{Abstract}
Symmetries play a major role in
physics, in particular since the  work by E.~Noether and H.~Weyl in the
first half of last century. Herein, we briefly
review their role by recalling how symmetry changes allow to
conceptually move from classical to relativistic and quantum physics.
We then introduce our ongoing theoretical analysis in biology and show
that symmetries play a radically different role in this discipline,
when compared to those in current physics. By this comparison, we
stress that symmetries must be understood in relation to conservation
and stability properties, as represented in the
\emph{theories}. We posit that
the dynamics of biological organisms, in their various levels of
organization, are not ``just'' processes, but
permanent (extended, in our terminology) critical transitions and,
thus, symmetry changes. Within the limits of a relative structural
stability (or interval of viability), variability is at the core of
these transitions. 

\paragraph{Keywords: } symmetries, systems biology, critical transitions,
levels of organization, hidden variables, coherent structures, downward
causation.

\section{Introduction and summary  }

A synthetic understanding of the notion of organism requires drawing
strong correlations between different levels of organization as well as
between the global structure and the local phenomena within the
organism. These issues should govern any systemic view on biology. Here,
we sketch an approach in which the living state of matter is
interpreted as a permanent  ``transition'', conceived as an ongoing or
\emph{extended} and \emph{critical} transition. A large amount of
very relevant work pertaining to the Theories of Criticality in physics
has been successfully applied to biology (see below). The mathematical
core of these theories rests upon the idea that a ``phase transition,''
which can be either critical or not, may be described as a
\emph{point} along the line where the intended control parameter
runs. For example, the ferromagnetic / paramagnetic transition takes
place for a precise value of the temperature, the Curie temperature.
Mathematically, this is expressed by the ``pointwise'' value of this
temperature, i.e., one mathematical point in this parameter’s space.
When the temperature decreases and passes through that point, the
magnetic orientation organizes along one direction and magnetism
appears. When the temperature increases through that point, disorder
prevails and magnetism disappears. A (phase) transition is critical
when some observables, or their first or second derivatives, diverge.
This corresponds to the appearance of a ``coherent structure'', that is
to say space and/or time correlations at all scales, which at the
transition point give a ``global'' aspect to the new physical object.
These ideas are relevant to the analysis of biological organisms. 

In contrast to known critical transitions in physics, biological
entities should not be analyzed just as transient over a point of a
phase change; instead, they permanently sustain criticality over a
non-zero interval and this with respect to many control parameters
(time, temperature, pressure). This represents a crucial change of
perspective. First, the mathematical tools used in physics for the
analysis of criticality, i.e, the renormalization methods, essentially
use the pointwise nature of the critical transitions. Secondly,
\emph{symmetries} and \emph{symmetry breakings} radically change
when enlarging the mathematical locus of criticality from one point to
a non-zero interval. These symmetry changes make a key theoretical
difference with respect to the few cases in physics where the
transition seems extended (see footnote 10, below). Our approach may be
seen as a move from physics to biology by an analysis of the radically
different symmetries and symmetry breakings at play in their respective
theoretical frames. Thus, we will mostly focus on physical vs
biological criticality in terms of symmetries and then apply this
method to the analysis of the difference between physical and
biological ``objects'' as well as of physical vs biological
``trajectories''.

Living entities are not ``just'' processes, but something more:  they are
lasting, \emph{extended critical transitions}, always
transient toward a continually renewed structure. In general, physical
processes do not change fundamental symmetries:  to the contrary, they
are mostly meant to preserve them. Typically, conservation properties
(of energy, of momentum) are symmetries in the equations of movement.
Critical transitions are an exception to the preservation of symmetries
in physics; their ``extension'' radically changes the understanding of
what biological processes are. This perspective also proposes a
possible way of overcoming a key issue in the analysis of the
complexity of the living state of matter. As for the construction of
physico-mathematical or computational models, it is difficult to take
the global structure of an organism into consideration, with its
correlations between all levels of organization and in all lengths,
including the many forms of integration and regulation. Thus, the
complexity of the living unity is often modeled by the stacking of
many but \emph{simple }elementary processes. Typically, these formal
systems deal with many observables and parameters. Since the framework
is classical in a physical sense, these variables are local, i.e. they
depend on pointwise values of the intended phase space. Instead,
conceptual and mathematical dependencies in biology should be dealt
with as ``global'' ones, where variables may depend on systemic or
\emph{non-local} effects. In physics, these dependencies are a
relevant aspect of critical transitions, and they are even more so in
biology, where criticality is extended.

\subsection{Hidden variables in biology? }

In classical and relativistic physics, once the suitable ``phase
space'' and the equations that mathematically determine
the system are given, the knowledge of the pointwise position-momentum
of the intended object of analysis allows to describe \emph{in
principle} the subsequent dynamics. This is ``in
principle'' since physical measurement, which is always
approximated, may produce the phenomenon of \emph{deterministic
unpredictability}, in particular in the presence of non-linear
mathematical determination\footnote{ More generally,
unpredictability may appear when the dynamics is determined by an
evolution function or equations that mathematically represent ``rich''
interactions. Non-linearity is a possible mathematical way to express
them.}. Moreover, not all ``forces'' in the game may be known and there
may be ``hidden variables'' (like the frictions along the trajectory of
bouncing dice). Yet, these theories are deterministic and, once all
pertinent variables and forces are assumed to be known, it is the
\emph{epistemic} lack of knowledge which yields classical randomness.
\emph{Per se}, a dice follows a ``geodesic''. This is a unique, optimal
and  ``critical'' path, completely determined by the Hamiltonian and may
be computed as an optimum of a Lagrangian
functional.\footnote{ These are mathematical
operators, that is, functions acting on functions that contain all
known physical information concerning the energy state of the
system.}. This very beautiful paradigm, which may be summarized as the
``geodesic principle'', may be further grounded on \emph{symmetries} by
an analysis of conservation principles (see \nptextcite{bailly2011} for
a recent synthesis and references).

In order to compare this situation with other fields of physics and
subsequently to biology, we refer to the pointwise or local nature of
the mathematical variables. Cantorian (and Euclidian) points are
\emph{limit} conceptual constructions; that is, they are the limit of
a physical access to space and time by an always approximated
measurement, i.e., an ``arbitrarily small'' interval. Yet,
their perfect theoretical ``locality'' makes all classical dynamics
intelligible (in principle). So, if something is unknown, one expects
that by adding enough observables and/or more variables with definite
values at any given time, one could increase knowledge, since the
values of these observables are intrinsic and independent of the
context.

The situation is rather different in Quantum Mechanics. The
simultaneous, perfect, pointwise knowledge of position \emph{and}
momentum (or energy \emph{and} time) are, in
principle, forbidden because indeterminacy is
intrinsic to the theory. Moreover, suppose that two quanta interact and
form one system and that they later separate in space. Then acquiring
knowledge regarding an observable quantity by performing a measurement
on one of these quanta produces an instantaneous knowledge of the value
of the measurement made on the other, i.e., the two quanta are
``entangled'' \parencite{EPR}. These features of the theory have
several consequences:  for instance, variables cannot always be
associated to separated points and quantum randomness is intrinsic
(under the form of Schrödinger equation, the ``determination'' gives the
\emph{probability} to obtain a value by measurement). Within this
theoretical framework, quantum randomness differs from the classical
one:  two interacting dice which later separate obeying independent
statistics, while the probability values of an observable of two
previously interacting quanta are correlated. This is the so called
``violation of Bell inequalities'', which has been empirically verified
repeatedly since the experiments described in \textcite{Aspect}.
Quantum entanglement requires considering some phenomena as being
``non-local'' and unseparable by any physical measurement
(``non-separability'').

Since the ’30s, some have found this situation unsatisfactory and have
searched for ``hidden variables'' like in the epistemic approach to
randomness and determination of classical and relativistic physics. The
idea is that these hidden variables corresponding to quantum mechanical
observables have definite (pointwise/local) values at any given time,
and that the values of those variables are intrinsic and independent of
the device used to measure them. A robust result has instead shown that
these assumptions contradict the fundamental fact that quantum
mechanical observables need not be commutative \parencite{Kochen}. Moreover, even when assuming the existence of, or the need for,
hidden variables, these would be ``non-local'' and thus, far from the
pointwise/local dependence of set-theoretic variables.

The difference between the classical and quantum frameworks has the
following consequence:  quantum systems may have a proper systemic unity
for at least two reasons. Conjugated observables (position and
momentum) are ``linked'' by joint indetermination. Entangled quanta
remain a ``system'', in the sense of their non-separability by
measurement\footnote{ Superposition should also be
mentioned, see \textcite{Silverman}.}.

Can this perspective help us in biology?  On technical grounds, surely
not, or rather not yet. Perhaps, ``entangled molecular phenomena'' or
``tunnel effects \dots\ in the brain'' may clarify fundamental issues in the
future. However, theoretical ideas in Quantum Mechanics may at least
inspire our attempts in system biology, in particular by considering
the methodological role of symmetries and symmetry breakings in this
area of physics.

A living organism is a system. And entanglement, non locality,
non-separability, superposition, whatever these concepts may mean in
biology, may present themselves both at each specific level of
organization and in the interactions between levels of organization.
Physiological interactions among molecules, cells, tissues, organs~ do
not simply sum each other up:  they are ``entangled'', ``non-local'',
``non-separable'' \dots\ they are ``superposed'' (see examples described by
 \nptextcite{noble2006,soto2008}). Thus, the theoretical and mathematical
approaches to biology cannot be based only on a continual enrichment of
``local'' views:  mathematical models cannot work just by assuming the
need for more and more variables (possibly hidden to the previous
models). A global view of the system and of its symmetries is required.
In this context, the differences in symmetries and their breakings will
help in clarifying and facilitating the passage from physics to
biology.

\section{Symmetry and objectivation in physics}

In Physics, objectivity is obtained by the co-constitutive use of
experiments and mathematized theories. So far, however, there is little
mathematics for a ``theory of the biological organisms'' despite the
large amount of data collected and of theories proposed within specific
levels of organization. These include the geometric analysis of the
fractal structures of lungs, of vascular systems, of various plant
organs, of networks of neural cells, of tumor shapes, to name but a
few. To make further progress towards mathematizing theories in
biology, in particular towards theories of the ``living object'' or of
the organism as a system, it would help first to understand how such a
feat was achieved in physics. Physical theories have very general
characteristics in their constitution of objectivity, and in particular
in their relationship with mathematics. In order to define space and
time, as well as to describe physical objects, physicists ultimately
use the notion of symmetry. Physical symmetries are the transformations
that do not change the intended physical aspects of a system in a
theory. As we shall see, they allow to define these aspects in a
non-arbitrary way.

Galileo’s theory provides a simple and historical example of this role
of symmetries.  For scholastic physics, the speed at which a body falls
is proportional to the space traveled. Galileo instead proposed that it
is proportional to the time of the fall and that it is independent of
the nature (including the mass) of the empirical object considered
(Galileo’ law of gravitation). This idea  together with the ``principle
of inertia'' has been a starting point for the constitution of
\emph{space} and \emph{time} in classical physics. More precisely,
as a consequence of the analysis of inertia and gravitation, the
geometry of space and time was later described by the Galilean
group\footnote{ Symmetries form a set of
transformations that have a group structure; that is, two symmetries
applied successively yield a symmetry and a symmetry can be inverted.
Galileo’s group is the group of transformations that allows to
transform a Galilean space-time reference system into another. It is
interesting to notice that Galileo measured time by heartbeat, a
biological rhythm; the subsequent theoretical and more ``physical''
measurement of time were
precisely provided by classical mechanics, his invention.}.

A  change of this symmetry group, for example by adopting the Poincaré
group\footnote{ The symmetry group of a Euclidean
space is the Euclidean group of automorphisms, while Poincaré’s group
corresponds to the automorphisms defining Minkowski’s spaces.}, can
lead to a very different physical situation, that of special relativity
involving massive conceptual and physical changes. The ``principle of
relativity'' states that the fundamental laws of physics do not depend
on the reference system; they are actually obtained as invariants with
respect to the change of reference system. A specific speed (the speed
of light in the void) appears in the equations of electromagnetism.
Einstein modified Galileo’s group in order to transform this speed into
an invariant of mechanics, which turned time-simultaneity into a
relative notion. 

As a result of the role and implications of symmetries, most
contemporary physical challenges lead to the search for the right
symmetries and symmetry changes, such as the work aiming at the
unification of relativistic and quantum theories.  In moving from
physics to biology we suggest here to apply a similar approach
(symmetry changes).

Since the 1920s, due to Noether’s theorems, symmetries lead to the
mathematical intelligibility of key physical invariant quantities. For
example, symmetries by time translations are associated with
energy-conservation, and symmetries by space rotations are associated
with the conservation of angular momentum. Thus, conservation laws and
symmetries are in a profound mathematical relation. Consequently, the
various \emph{properties} that define an object (mass, charge,~etc.)
or its \emph{states} (energy, momentum, angular momentum,~etc.) are
associated to specific symmetries which allow these quantities to be
defined. Depending on the theory adopted, this conceptualization
allowed to understand why certain quantities are conserved or not:  for
example, there is no local energy conservation in general relativity.
This explicit reference to the theory adopted is required in order to
produce ``scientific objectivity'', \emph{independently}
of the arbitrary choices made by the observer, such as, the choice of
time origin, the unit of measurement, etc, but \emph{relatively} to
the intended theory. Thus, we say that symmetries provide ``objective
determinations'' in physics \parencite{bailly2011}.

The symmetries that define physical properties allow us to understand
the physical object as \emph{generic}, which means that any two
objects that have the same properties can be considered as physically
\emph{identical}; in a sense, they are symmetric or invariant
(interchangeable) in experiments and in pertinent mathematical
framework (typically, the equations describing movement). For example,
for Galileo, all objects behave the same way in the case of free fall,
regardless of their nature. Moreover, symmetries allow the use of the
\emph{geodesic principle,} whereby the local determination of
trajectories leads to the determination of the full trajectory of
physical objects through conservation laws. For example, the local
conservation of the ``tangent'' (the momentum) of movement, typically
yields the global ``optimal'' behavior of the moving object; that is, it
goes along a geodesic. Thus, in classical or relativistic mechanics, a
trajectory is unique and fully deterministic (formally determined). In
quantum mechanics the evolution of the state or wave function (roughly,
a \emph{probability distribution}) is fully deterministic as well –
and determined by Schrödinger’s equation – while measurement follows
this probability distribution (and here appears the indeterministic
nature of quantum mechanics). In conclusion, by symmetries, the
trajectory of a generic classical or quantum physical ``object''
corresponds to a critical path:  physical trajectories are
\emph{specific}. 

 To better understand the problem of \emph{general} mathematical
theorizing in biology, let’s further analyze how, in physics, a
concrete problem is turned into robust models and mathematics. To begin
with, physicists try to choose the right theoretical framework and the
relevant physical quantities (properties and states) which are
constituted by proper symmetries. As a result, typically, a
mathematical framework is obtained, where one can consider a generic
object; in classical mechanics, a pointwise object of mass~$m$,
speed~$v$ and position~$x$, where these quantities are
generic. Now, a generic object will follow a specific trajectory
determined by its invariants obtained by calculus. A measurement is
then made on the experimental object to determine the quantities
necessary to specify where this object is in this mathematical
framework, namely, what is its mass, initial position and speed. And
finally, what specific trajectory will the object follow \dots\  at least
approximately. In classical or relativistic physics, to a specific
measurement will correspond generic objects localized near the
measurement due to the limited precision of this measurement. This
value may have, in principle, an arbitrary high precision. In quantum
mechanics, as we recalled above, the equational determination
(Schroedinger’s equation) yields the dynamics of a probability
law\footnote{In quantum physics, ``objects'' do not
follow trajectories in ordinary space-time, but they do it in a
suitable, very abstract space, a Hilbert space (a space of mathematical
functions); what ``evolves'' is a probability distribution.}.

In classical dynamics, we face a well-known problem:  the specific
trajectories can either stay close or diverge very rapidly. The linear
situation corresponds to the first case, whereas the second situation
is called ``sensitive to initial conditions'' (or chaotic, according to
various definitions). Note that even the latter situation leads to the
definition of new invariants associated to the dynamics:  in other
words, the attractors that have a precise geometrical structure. In
both cases, these trajectories have robust properties with respect to
the measurement. In quantum physics, the situation is more complex
because the measurement is not deterministic. Yet, when approximations
on the state function are performed, it leads to usually stable. robust
statistics. In all cases, ``robust'' means invariant or approximately
invariant in a definite mathematical sense, as concerns the measurement
of states and properties of generic objects along specific
trajectories. Thus, we can finally say that generic objects, which lead
to a specific measurement, \emph{behave} in the same way or
approximately so. Notice that this property of robustness, allowed by
the genericity of the object, is mandatory for the whole framework to
be relevant. We insist that both genericity for objects and specificity
for trajectories (geodesics) are mathematically understood in terms of
symmetries. 

In conclusion, in the broadest sense, symmetries are at the foundation
of physics, allowing objective definitions of space and time and the
constitution of objects and trajectories. In their genericity, these
objects follow specific trajectories associated with invariants that
are robust with respect to measurement.

\section{  Symmetry breakings and criticality in physics }

The physics of criticality is a relatively novel discipline which
analyzes, typically by the renormalization techniques, some peculiar
phase transitions, i.e., state changes (see \nptextcite{toulouse1977introduction,Binney}). This theoretical framework has also been applied
to a possible understanding of life phenomena (see for example, \nptextcite{Bak88,Jensen}, as for ``self-organized criticality''; or,
\nptextcite{Kauffman93}, as for criticality in networks). We will next move
towards biology through a different insight into the symmetries in
criticality.

Since symmetries are at the core of the definition of the physical
objects by their properties and states, a \emph{symmetry change}
(that is, the breaking of some symmetries and the formation of new
ones) means a qualitative change of the object considered, or a change
of physical object, understood as co-constituted by theory and
empiricity. For example, a research project in cosmology is to consider
a single force to have existed in the universe right after the big
bang. Then, the four fundamental forces may have appeared by successive
symmetry breakings, whereby some transformations, which were
symmetries\footnote{The Higgs mechanism is an example
of this phenomenon; in this case, the symmetry breaking in the abstract
electroweak space leads in particular to different masses of bosons and
as a consequence to a very short range for weak interaction and a long
range for electromagnetism.}, did not preserve the object invariance
anymore. In other words, with the cooling of the universe, the system
moved to a smaller symmetry group. Closer to the scale of biology,
materials like water or iron were able to show different properties in
different situations. Depending on the temperature and pressure, water
may be a solid, a liquid,  or a gas. When liquid, there is no
privileged direction (the system is isotropic, that is to say symmetric
by rotations), whereas ice has a crystalline structure with spatially
periodic patterns. This implies that the system is no longer symmetric
by continuous rotations:  it has a few privileged directions determined
by its crystalline structure and a smaller symmetry group. Similarly,
iron can have paramagnetic behavior (the system is not) or
ferromagnetic behavior (it is magnetized). In most cases, one can
distinguish a more disordered phase at high temperature, where entropy
dominates, and a more ordered phase, where energy dominates. These
situations can be characterized by an \emph{order parameter} which is
$0$ in the disordered phase and different from $0$ in the ordered
phase\footnote{ Here, order means low entropy (or
less symmetries) and disorder means high entropy (and more symmetries,
when symmetries are computed in terms of ``microstates'').}.

Now, in physics, the change of state, or \emph{phase transition},
occurs always mathematically at a point of the parameters’ space. This
point, called the \emph{critical point}, is intuitively associated
with a sudden change of behaviour due to a change of symmetry, and
ultimately to singularities of the state functions (for example, the
order parameter is non-analytical because it goes from a
\emph{constant} 0 to a finite quantity, \emph{by a finite change}).
More technically, the \emph{critical point} represents a singularity
in the partition function describing the
system\footnote{ This function is non-analytical at
the critical point, which means that the usual Taylor expansions,
linearizations or higher order approximations do not actually provide
an increasing approximation.}. In the case of iron’s
paramagnetic-ferromagnetic transition, this allows to deduce the
divergence of some physical observables, such as magnetic
susceptibility. It should be remembered that this notion of
\emph{singularity,} which is associated with infinite quantities at
the critical \emph{point}, is a core notion for physical criticality.

This peculiar situation leads to a very characteristic behaviour at the
critical point \parencite{Jensen}:  
\begin{enumerate}
 \item Correlation length tends to infinity, and follow a power law, as
for continuous phase transitions (i.e., for a vector~$x$ and an
observable~$N$, if we note by~$<.>_r$ the average over point~$r$ in
space, then  $< N(r + x)N(r)>_r-<N(r)>_r^2\sim \|x\|^\alpha$. This is
associated with fluctuations at all scales leading in particular to the
failure of mean field approaches. Following this approach, the value of
an observable at a point is given by the mean value in its
neighbourhood or, more precisely, its mathematical distribution is
uniform. 
 \item Critical slow down:  the time of return to equilibrium of the
system after a perturbation tends to infinity \parencite{suzuki1982critical,tredicce2004critical}. 
 \item Scale invariance:  the system has the same behavior at each scale.
This property leads to fractal geometry and means that the system has a
specific symmetry (scale invariance itself). 

 \item The determination of the system is global and no longer local. 
\end{enumerate}

These properties are the key motivations for the biological interest of
this part of physics. The global ``coherence structure'' that is often
formed at critical transitions provides a possible understanding, or at
least, an analogy for the unity of an organism (in current terminology,
its ``global determination or causation''). Also, power laws, so frequent
in biology, are ubiquitous in critical phenomena. They are
mathematically well-behaved functions (e. g. $f(x) =
x^\alpha$) with respect to the change of scale
[typically, ${\lambda}$ is the scale change in  $f(\lambda x) =
\lambda^\alpha f(x) =
\lambda^\alpha x^\alpha$, 
a power law in $\alpha$], and they yield \emph{scale symmetries.
}In our example, scale change just multiplies the function $f$ by a
constant $\lambda^\alpha$. Now, a power law
depends on a quantity without physical dimension ($\alpha$ in the
notation above). These quantities involved in critical transitions are
called \emph{critical exponents} and describe how the change of scale
occurs. In our terminology, they describe the  properties due to the
objective determination of a phase transition because they are the
invariants associated with the scale symmetry.

Specific analytical methods, called renormalization methods, are used to
find these quantities \parencite{delamotte2004hint}. These methods consist in
analyzing how scale changes transform a model representing the system,
and this analysis is made ``asymptotically'' toward large
scales. One may deduce the critical exponents from the mathematical
operator representing the change of scale. The key point is that a
variety of models ultimately lead to the same quantities, which means
that they have the same behavior at macroscopic scales. Thus, they can
be grouped in so-called \emph{universality classes}. This analytical
feature is confirmed empirically, both by the robustness of its results
for a given critical point and more stunningly by the fact that very
different physical systems happen to undergo the same sort of phase
transitions; that is, they are associated with the same critical
exponents, thus with the same symmetries. Finally, there exist
fluctuations at all scales, which means, in particular, that small
perturbations can lead to very large fluctuations.

To conclude, the transition through a specific point of the parameters’
space, i.e., a transition between two very different kinds of behavior
is associated in physics to a change of symmetries. At this point, the
system has very peculiar properties and symmetries. Symmetries by
dilation (by a coefficient $\lambda$ as above) yield a scale
invariance. This latter invariance is associated to a global
determination of the system and the formation of a ``structure of
coherence''. As observed above, this allows to describe a global
determination of local phenomena and a unity that by-passes the idea of
understanding the global complexity as the sum of many local behaviors
by adding more and more local, possibly hidden, variables. For some
physical phenomena this theoretical framework presents peculiar and
very relevant forms of ``systemic unity''.

\section{  Symmetry breaking and the biological object:  extended criticality}

We have presented a picture of the situation in physics, but what about
biology?  We need to propose one or several specific frameworks
relevant to the unity and coherence of biological entities, because, to
our knowledge, there are no formalized theories of the ``organism''. To
do so, it may be worthwhile to look at the symmetries which may be
involved in biological theorizing. Here, the concept of symmetry is
used in a more fundamental context than when used, for example, for
``bauplans'', the latter being the main biological research subjet where
the concept is explicitly applied. In physics, one mostly deals with
\emph{fundamental} or \emph{theoretical} symmetries as typically
given by the equations. For example, the already mentioned fundamental
principle of energy conservation corresponds to a time translation
symmetry in the equations of movement.  This use of symmetries also
justifies the soundness of empirical results:  Galilean inertia is a
special case of conservation of energy and it may be empirically
verified. In biology, as in any science, a missing analysis of
invariants may give unreliable results and data. For example, early
measurements of membrane surfaces gave very different results, since
their measure is not a scale invariant property:  as in fractal
structures, it depends on the scale of
observation\footnote{In \textcite{weibel1994}, another
``historical'' example is given as for the different results that are
obtained according to different experimental scales (microscope
magnifications). One team evaluated the surface density of the liver’s
endoplasmic reticulum at \SI[per-mode=symbol]{5.7}{\square\metre\per\cubic\centi\metre}
the other at \SI[per-mode=symbol]{10.9}{\square\metre\per\cubic\centi\metre}
(!). }. In other words, in physics, both the
generality of equations and the very objectivity of measures depend on
theoretical symmetries and their breakings, such as scale invariants
and scale dependencies.

As mentioned above, critical transitions in physics are mathematically
analyzed as isolated points\footnote{The
Kosterlitz-Thouless transition in statistical physics presents a
marginally critical interval; that is, it is a limit case between
critical and not critical. It presents correlations at all scales, as
critical features, but with no symmetry changes. Thus, this particular
situation is not a counter-example to our statement (the essentially
pointwise nature of the proper physical transitions),
in view of a lack of symmetry changes that are essential to our notion
of extended criticality.}. In our approach to biological processes as
``\emph{extended} critical transitions'', ``extended'' means that
\emph{every point} of the evolution/development space is near a
critical point. More technically, the critical points form a
dense\footnote{ Here, dense means that for every
small volume of the intended phase space being considered, there is a
critical point in such volume.} subset of the multidimensional space
of viability for the biological process. Thus, criticality is extended
to the space of all pertinent parameters and observables (or phase
space), within the limits of viability (tolerated temperature, pressure
and time range, or whatever other parameter, say for a given animal), see
 \textcite{bailly1993,bailly2008,bailly2011}. In
terms of symmetries, such a situation implies that biological objects
(cells, multicellular organisms, species) are in a \emph{continual
transition between different symmetry groups}; that is, they are in
transition between different phases, according to the language of
condensed matter\footnote{The dense set of symmetry groups may be
potentially infinite, but, of course, an organism (or a species)
explores only finitely many of them in its life span, and only viable
ones. }. These phases swiftly shift between different critical points
and between different \emph{physical determinations} through symmetry
changes.

Our perspective provides an approach concerning the mathematical nature
of biological objects as a \emph{limit} or asymptotic case of
physical states:  the latter may yield the dense structure we attribute
to extended criticality only by an asymptotic accumulation of critical
points in a non-trivial interval of viability --- a situation not
considered by current physical theories. In a sense, it is the very
principles grounding physical theories that we are modifying through an
``actual'' limit. Thus, a biological object is mathematically and
fundamentally different from a physical object because it may be
characterized in terms of partial but continual changes of symmetry
within an interval of viability, as an extended locus of critical
transitions. In particular, this mathematical view of ``partial
preservation through symmetry changes'' is a way to characterize the
joint dynamics of \emph{structural stability} and
\emph{variability} proper to life. We thus consider this
characterization as a tool for the mathematical intelligibility of
fundamental biological principles: the  global/structural stability is crucially associated 
with variability.

A first consequence of these permanent symmetry changes is that there
are very few invariants in biology.  Mathematically, invariants depend
on stable symmetries. Structural stability in biology, thus, should be
understood more in terms of \emph{correlations of symmetries within
an interval of the extended critical transition,} rather than on their
identical preservation. It is clear that the \emph{bauplan} and a few
more properties may be ``identically'' preserved. Yet, in biology,
theoretical\emph{ }invariants are continually broken by these
symmetry changes. A biological object (a cell, a multicellular
organism, a species) continually changes symmetries, with respect to
all control parameters, including time. Each mitosis is a symmetry
change because the two new cells are not identical. This variability,
under the mathematical form of symmetry breaking and constitution of
new symmetries, is essential both for evolution and embryogenesis. The
interval of criticality is then the ``space of viability'' or locus of
the possible structural stability. 

The changes of symmetries in the dense interval of criticality, which
provide a mathematical understanding of biological variability, are a
major challenge for theorizing. As a matter of fact, we are
accustomed to the theoretical stability warranted by the mathematical
invariants at the core of physics. These invariants are the result of
symmetries in the mathematical (equational) determination of the
physical object. This lack of invariants and symmetries corresponds to
the difficulties in finding equational determinations in
biology\footnote{In a rather naive way, some say
this by observing that any (mathematized) theory in biology has a
``counterexample''. This instability of the
determination goes together with the ``structural
stability'' of biological entities. This is largely due to
the stabilizing role of integration and regulation effects between
different levels of organization. The mathematics of extended
criticality and of variants of the renormalization methods are yet to
be developed.}.

As a further consequence of our approach, phylogenetic or ontogenetic
trajectories cannot be defined by the geodesic principle, since they
are not determined by invariants and their associated symmetries. These
latter are continually changing in a relatively minor but extended way.

Biology may be considered to be in an opposite situation with respect to
physics:  in contrast to physics, in biology, \emph{trajectories
}are\emph{ generic }whereas\emph{ objects }are \emph{
specific} \parencite{bailly2011}. That is, a rat, a monkey
or an elephant are the \emph{specific} results of \emph{possible}
(generic) evolutionary trajectories of a common mammal ancestor --- or
each of these individuals is \emph{specific}. They respectively are
the result of a unique constitutive history, yet a possible or
\emph{generic} one \parencite{bailly1993,bailly2011}.

The evolutionary or ontogenetic trajectory of a cell, a multicellular
organism or a species is just a \emph{possible} or
\emph{compatible} path within the ecosystem. The genericity of the
biological trajectories implies that, in contrast to what is common in
physics, we cannot mathematically and \emph{a priori} determine the
ontogenetic and phylogenetic trajectory of a living entity be it an
individual or a species. In other words, in biology, we should consider
\emph{generic} trajectories (or possible paths) whose only
constraints are to remain compatible with the survival of the intended
biological system. Thus, phylogenesis and embryogenesis are
\emph{possible} paths subject to various constraints, including of
course the inherited structure of the \textsc{dna}, of the cell and the
ecosystem. The \emph{specificity} of the biological object, instead,
is the result of critical points and of symmetry \emph{changes} of
the system considered \emph{along its past history} (evolutive and
ontogenetic). These constitute the specific ``properties'' of this
object, which allow to define it. A rat, a monkey or an elephant or
their species are \emph{specific} and cannot be interchanged either
as individuals nor as species. A living entity is the result of its
history and cannot be defined ``generically'' in terms of invariants and
symmetries as it is done for physical objects.

This situation has a particular meaning when we consider time
translation and time reversal symmetries. In physics, time symmetries
correspond to the maintaining of the system’s invariant quantities that
define the geodesics, as for example, conservation of energy. In
biology both symmetries are broken. In particular, evolutionary and
ontogenetic paths are both irreversible and non-iteratable; there is no
way to identically ``rewind'' nor ``restart'' evolution or
ontogenesis. This corresponds to the breaking of time translation and
reversal symmetries. In particular, this lack of time symmetries is
associated with the process of \emph{individuation}, understood here
as the specificity of cells, organisms and species (as much as this
latter notion is well defined). It is crucial to understand that time
plays a key role in this framework, since the \emph{history} of all
the changes in symmetry are not reducible to a specific trajectory in a
given space of the dynamics. Thus, 

{\centering
\emph{The sequence of symmetry changes defines the historical
contingency of a living object’s phylogenetic or ontogenetic
trajectory}.
\par}

Biological processes are more ``history based'' than physical processes.
Usual physical processes preserve invariants, whereas extended critical
transitions are a permanent reconstruction of organization and
symmetries, i.e., of invariants. This situation also points to a lack
of symmetry by permutation. For example, even in a clonal population of
bacteria, different bacteria are not generic, because they are in
general not interchangeable, i.e., they cannot be permuted. This allows
to understand biological variability in a deeper way than the usual
Gaussian (or combination of Gaussians) as random distribution of a set
of observables. Now, let us consider organs (and organelles). Some
organs have a functional role that can be expressed in a physical
framework, particularly as far as energy transfer is concerned. This
functional role can lead to restrictions on the variability of the
cells that constitute the organ, while the same could be said for
individual organisms in populations. At least for certains aspects of
their behaviour and on average, these restrictions make cells behave 
symmetrically. In other words, those cells behave, in part and
approximately, like generic objects with specific trajectory
(geodesics). They may be interchangeable, like physical objects. 

The simple case of cells secreting a protein such as erythropoietin
(\textsc{epo}) under specific conditions indicate that on average, a sufficient
amount of the protein must be produced, independently of the individual
contribution of each cell (which become ``relatively'' generic). Since
the result of these cells’ production is additive (linear), its
regulation does not need to be sharp. Even if some cells do not produce
\textsc{epo} there is no functional problem as long as a sufficient quantity of
this protein is secreted at the tissue level. However, when cells
contribute to a non-linear framework as part of an organ, the
regulation may need to be sharper. This is the case, for example, for
neuronal networks or for cell proliferation where non-linear effects
may be very important. In the latter case, regulation by the tissue and
the organism seems to hold back pathological developments, like cancer,
see \textcite{Society}. This point of view can possibly be
generalized in order to understand the robustness of development. 

The role of physical processes in shaping organs is crucial; for
example, exchanges of energy (or matter) force/determine the optimal
(geodesic) fractal structure of lungs and vascular systems. Organs in
an organism may even be replaced by man-made artifacts (as for kidneys,
heart, limbs, etc.). As biological entities, organisms and even cells
are specific or, at most, weakly generic given that they can be
interchanged only within a given population or tissue and occasionally. 
In general, they are not generic, and by their specificity they cannot
be replaced by an artifact --- structurally.

In summary, in critical transitions one may consider variables depending
on global processes because of the formation of coherent structures.
For example, there may be functional dependencies on a network of
interactions, which cannot be split into a sum of many local
dependencies (local variables). Thus, the search for more variables
would not take into account this fundamental property of biological
systems, considered as extended critical transitions. Moreover,
symmetries in physics allow to define generic objects which follow
specific trajectories (the latter allowing to find invariants in terms
of symmetries, which are robust regarding measurement). On the
contrary, in biology, the continual symmetry changes lead to generic
trajectories that remain compatible with the survival of the system.
The generic/specific duality with respect to physics helped us
understand this key issue, in relation to extended criticality --- which
is a form of ``relatively stable instability.'' In other words, this is
stability under changes of symmetries in an interval of viability. In a
sense, the biological object is also defined by its symmetries but in a
very different way:  it is the \emph{specific} result of a history,
where its dynamics is punctuated by symmetry changes. This makes it
``historical'' and \emph{contingent}.

\section{Additional characteristics of extended criticality}

In physics, criticality implies more than a pointwise symmetry change;
that is, it requires a change on a mathematical point, as it leads to
peculiar behaviors that are relevant to biology. The first of these
properties is that criticality implies a global determination, instead
of a simply local one. More precisely, the singularities involved in
criticality lead to a change of the level of organization in a very
strong sense. Also in physics, in view of the mathematical divergence
of some observables, the singularities break the ability of the ``down
level'' to provide a causal account of the phenomena and they lead to
the need for a ``top level'' to overcome this difficulty. In mathematical
physics, this upper level can be found in the renormalization operator
(it is the abstract level of \emph{changing scale}). In biology,
instead, the upper level is the functional unity of an organism. As a
result, the existence of different levels of organization is a
component of our notion of extended critical transition. ``Downward
causation'' may find the right frame of analysis in this theoretical
context. 

The permanent reconstruction of these levels of organization is
mathematically represented by the density of the critical points and by
the continual change of determination (symmetry change) in the passage
between these points within the interval of extended criticality. 

The second property is the presence of power laws which seem to be
ubiquitous in biology.  They appear regularly especially when
regulation is concerned, such as  in
cardiac rhythms \parencites{Makowiec2006,pikkujamsa1999cardiac}, blood
cell number regulation \parencite{Perazzo00}, blood pressure \parencite{wagner1996chaos}, in brain activities \parencite{GerhardWerner07}, sensory cells
\parencite{SebastienCamalet99}, mitochondrial networks \parencite{aon2004percolation}, in ecology \parencite{Sole99} and
gene networks \parencite{IlyaShmulevich05,nykter2008gene}. 

Extended critical transitions also concern the relevant lengths of local
and global exchanges, the temporalities mobilized for such exchanges
and biological rhythms. To summarize, the extended critical situation
has at least the following characteristics \parencite{bailly2008,bailly2011}: 
\begin{enumerate}
 \item A spatial volume enclosed within a semi-permeable membrane; 
  \item Correlation lengths of the order of magnitude of the greatest
length of the above referred volume;
   \item A metabolic activity that is far from equilibrium and
irreversible, involving exchanges of energy, of matter and of entropy
with the environment, as well as the production of entropy due to all
these irreversible processes, see \textcite{bailly2009}; 
    \item An anatomo-functional structuralization into levels of
organization that can be autonomous but also coupled to each other.
They are ``entangled'' in the sense defined by \textcite{bailly2009,soto2008}. These
levels are likely to be distinguished by the existence of fractal
geometries (membranous or arborescent), where the fractal geometries
can be considered as the trace (or a ``model'') of effective passages to
the infinite limit of an intensive magnitude of the system (for
example, local exchanges of energy\footnote{The
fractal dimension of some organs may be calculated by optimizing the
purely physical exchanges within the intended topological dimension
(for example, the maximization, within a volume, of surfaces for lungs,
or of volumes for the vascular system, \nptextcite{west1997}), and it may be
subjected to constraints in terms
of stericity and homogeneity, as in the cases mentioned (lung, vascular
system, kidney, etc).}). The different levels of organization induce,
and are a consequence of, the alternation of ``organs'' and ``organisms'',
such as organelles in cells, which, in turn, make up the organs in
multicellular organisms. Organisms stay in
an extended critical transition, while organs are partially ``optimally shaped'' by
the exchange of physical energy and matter. For example, fractal
geometries essentially manifest in organs that are also the privileged
loci of endogenous rhythms (see below). Correlation lengths are
manifested both \emph{in} and \emph{between} these
levels\footnote{ The term ``entanglement'' in \textcite{soto2008} does not correspond, of course, to the physical meaning of
``quantum entanglement'' as expressed by Schrödinger’s treatment of the
state function and the inseparability of quantum measure, yet it may be
appropriate because there is no
way to isolate one of the organs mentioned above (e.g. put a brain in a
flowerpot) and perform any reasonable physiological measure on it.}.
Likewise, the various biological ``clocks'' are coupled, and in some
cases even synchronized, within and between these levels. 
\end{enumerate}

With the purpose of providing biological temporality with a structuring
of the mathematical type, we will consider two other aspects as being
specific to extended criticality.

\begin{itemize}
 \item The two-dimensionality of time, proposed in  \parencite{bailly2011b}:  
 \begin{enumerate}
  \item One dimension is classical and is parametrized according to the
line of real numbers limited by fertilization on one side, and death on
the other. This dimension is linked to the bio-physicochemical
evolution of the organism in relation to an environment. 
  \item The other dimension is compactified, i. e. it is parametrized on a
circle. This second dimension is linked to the organism’s endogenous
physiological rhythm that is manifested through \emph{numeric
quantities without dimension} such as the mean total number of
heartbeats and respirations during the lifetime of mammals. These are
the interesting interspecific invariants and they are ``pure'' numbers,
\emph{not frequencies} (they have no dimension; they are the ``total
number of \dots''). They  become frequencies (with the inverse of time as a
dimension), according to the average lifespan. The extra dimension is
needed exactly because the invariant phenomenon is not defined by a
period which has the dimension of time, but by this new invariant
observable. For example, on average, the identical (invariant) number
of total heartbeats give different frequencies according to the
different lifespans of an elephant or of a mouse.
 \end{enumerate}
Moreover, the temporality of extended criticality involve protention
(i.e. pre-conscious expectation) and retention (i. e. pre-conscious
memory)  \parencite{longo2011}, which seems to lead to a breaking of conservation of
information in cognition.

 \item The confinement within a volume of a parameter space (such as
temperature, pressure, etc) of $n$ dimensions of which $3$ are
spatial and $2$ temporal  and whose measure is different from $0$ (see
above).
\end{itemize}

\section{ Conclusion}

Since ancient Greece (Archimedes’ principle on equilibria) up to
Relativity Theory (and Noether’s and Weyl’s work) and Quantum Mechanics
(from Weyl’s groups to the time-charge-parity symmetry), symmetries
have provided a unified view of the principles of theoretical
intelligibility in physics. We claimed here that some major challenges
for the proposal of mathematical and theoretical ideas in biology
depend, in principle, on the very different roles that symmetries play
in biology when compared to physics. The unifying theoretical framework
in biology is neither associated to invariants nor to transformations
preserving invariants like in (mathematical/theoretical) physics. It
focuses, instead, on the permanent change of symmetries that
\emph{per se} modify the analysis of the internal and external
processes of life, both in ontogenesis and evolution. 

In a sense, variability may be considered as the main invariant of the
living state of matter. In order to explain it, we proposed to consider
the role played by local and global symmetry changes along extended
critical transitions. In extended criticality, dynamically changing
coherent structures as global entities provide an understanding of
variability within a global, extended stability.  The coherent
structure of critical phenomena also justifies the use of variables
depending on non-local effects. Thus, an explicitly systemic approach
may help in avoiding the accumulation of models and previously hidden
variables. In conclusion, the notion of extended criticality provides a
conceptual framework, to be further mathematized, where the dynamics of
symmetries and symmetry breakings provide a new, crucial role for
symmetries in biology with respect to physics.

\paragraph*{Aknowledgement: } We warmly thank the editors of the published version of this chapter for several and very close preliminary revisions of this conceptually difficult text.

\printbibliography[heading=bibintoc]
\end{document}